\documentclass[journal]{IEEEtran}

\usepackage{amsmath,amsfonts,amssymb}
\usepackage{graphicx}
\usepackage{array}
\usepackage{booktabs}
\usepackage{makecell}
\usepackage{diagbox}

\usepackage[caption=false,font=footnotesize]{subfig}

\usepackage{algorithm}
\usepackage{algorithmic}

\usepackage{stfloats}
\usepackage{textcomp}
\usepackage{verbatim}
\usepackage{url}
\usepackage{CJKutf8}

\usepackage{cite}

\usepackage[hidelinks]{hyperref}

\begin{document}

\title{Physics-Informed Feature Engineering 1D-CNN for Multilayer Cloud Detection from Geostationary Satellites}

\author{%
Fu~WANG,~\IEEEmembership{Senior Member ~IEEE,} Chi~YANG, Qifeng~LU,Ruixia~LIU,Xiaofei~Yang,Xiaofang~LIU,Bo~LI,Lin~CHEN%

        
\thanks{This work was supported in part by the National Natural Science Foundation of China under Grant U2242212 and 42471437, the National Key Research and Development Program of China under Grant 2022YFC3004102. (Corresponding author: Qifeng LU, Xiaofei Yang). 
Fu WANG, Qifeng LU, and Ruixia LIU are with the State Key Laboratory of Severe Weather Meteorological Science and Technology, CMA Earth System Modeling and Prediction Centre and Key Laboratory of Earth System Modeling and Prediction, China Meteorological Administration, Beijing, 100081, China (e-mail: wangfu@cma.cn, luqf@cma.cn, liurx@cma.cn).
Chi YANG and Xiaofang LIU are with the School of Computer Science and Engineering, Sichuan University of Science and Engineering, Yibin City, 643000, China (e-mail: yangchi\underline{~}001@foxmail.com,lxf1969@suse.edu.cn). 
Xiaofei Yang is with the School of Electronics and Communication Engineering, Guangzhou University, Guangzhou, 510006, China
Bo LI and Chen LIN are with the National Satellite Meteorological Center (NSMC) and the Innovation Center for FengYun Meteorological Satellite (FYSIC), China Meteorological Administration (CMA), Beijing, 100081, China (e-mail: libo@cma.cn, chenlin@cma.cn)}
\thanks{Manuscript received XX XX, 2025; revised XX XX, 2025.}}%

\markboth{Journal of \LaTeX\ Class Files,~Vol.~14, No.~8, August~2021}%
{Shell \MakeLowercase{\textit{et al.}}: A Sample Article Using IEEEtran.cls for IEEE Journals}

\maketitle %

\begin{abstract} 
Multilayer cloud detection from active-passive observation is vital for numerical weather prediction. In this study, channel selections derived from threshold-based algorithms are embedded as feature engineering priors into a 1D-CNN, and machine learning (ML) is used to learn latent physical relationships to simplify physical retrievals for operational deployment. The results reveal that the 1D-CNN attains a multilayer-cloud probability of detection (POD$_{\rm mul}$) of 0.620 and a false alarm rate (FAR$_{\rm mul}$) of 0.240, outperforming the conventional threshold algorithm (POD$_{\rm mul}$ = 0.558, FAR$_{\rm mul}$ = 0.369). Thereby demonstrating that prior physical knowledge derived from radiative transfer theory can serve as a feature engineering prior. Further experiments show that ML-revealed physical mechanisms can also enhance traditional algorithms. Replacing AGRI channel 12 (C12,centered at 10.8$\mu$m) with channel 13 (C13, centered at 12.0$\mu$m) increased POD$_{\rm mul}$ from 0.558 to 0.609 without materially affecting FAR$_{\rm mul}$. However, for AHI the substitution of the 11.2$\mu$m channel with 12.3$\mu$m yielded negligible improvement. Besides their spectral response functions (SRFs) mismatches, a primary contributing factor is the channels’ on-orbit radiometric stability. Hence, physics-informed, machine-learning methods appear promising for advancing remote-sensing AI, while sensor-specific characteristics must be considered during operational transfer.
\end{abstract}

\begin{IEEEkeywords}
Multilayer cloud detection, geostationary imagery, convolutional neural network, FY-4A/AGRI, Himawari-8/AHI. 
\end{IEEEkeywords}
\section{Introduction}
\IEEEPARstart{C}{louds} play a crucial role in the Earth's climate system, impacting both its radiative balance and hydrological cycles~\cite{dessler2010determination,trenberth2009earth}. Approximately 70\% of the Earth's surface is enveloped in clouds, 40\% of which are multilayered~\cite{yi_global_2025}. The multilayered clouds, particularly their diurnal variations, significantly contribute to the uncertainty of cloud climate effects~\cite{yin_diurnal_2017, noh_building_2024}. Precisely detecting and characterizing multilayer clouds, which can comprise various types of clouds at different altitudes, presents a significant challenge. Addressing this challenge is essential for enhancing numerical weather prediction (NWP) models~\cite{mace_description_2009}. Traditional satellite-based remote sensing techniques often struggle with multilayer cloud detection due to the overlapping layers, which impede the retrieval of cloud properties such as optical depth and phase~\cite{joiner2010detection}. 

Research on traditional multilayer cloud detection algorithms primarily utilizes polar-orbiting satellites equipped with imaging instruments, such as the Moderate Resolution Imaging Spectroradiometer (MODIS)~\cite{ackerman1998discriminating,platnick2003modis, wang_validation_2016}, the Visible Infrared Imaging Radiometer Suite (VIIRS)~\cite{wang2019multilayer} and the Polarization and Directionality of the Earth’s Reflectance (POLDER)~\cite{desmons_global_2017}. The MODIS Collection 6 (C6) multilayer cloud detection algorithm identifies multilayer cloud scenes with a cloud optical depth (COD) exceeding 4 using three sets of tests~\cite{platnick2016modis}. This algorithm integrates the band‐reflectance and brightness‐temperature‐difference (BTD) tests by Pavolonis and Heidinger~\cite{pavolonis2004daytime}, and exploits differences in cloud precipitable water content derived from shortwave infrared (SWIR) and longwave infrared (LWIR) channels. A strong agreement with CloudSat retrievals was observed, yielding a correlation coefficient of 0.834 for both single- and multilayer cloud detections~\cite{wind_multilayer_2010}. By leveraging contrasts in solar‐band reflectance, the VIIRS method simplifies the MODIS multi‐channel workflow and achieves roughly 60\% overall detection accuracy when validated against Cloud‐Aerosol Lidar with Orthogonal Polarization (CALIOP) observations onboard Cloud‐Aerosol Lidar and Infrared Pathfinder Satellite Observations (CALIPSO)~\cite{wang2019multilayer}. A similar multilayer-cloud detection algorithm has been extended to observations from the Advanced Geosynchronous Radiation Imager (AGRI) onboard the new generation of geostationary satellite FY-4A~\cite{wang_algorithm_2024}. Given their comparable band configuration, the algorithm can be readily adaptable to Himawari-8’s Advanced Himawari Imager (AHI) and GOES-16/17’s Advanced Baseline Imager (ABI)~\cite{wang_multi-layer_2024}.

Traditional methods rely on expert-tuned thresholds derived from benchmark statistics or in-situ measurements, but struggle to disentangle radiative signals from overlapping cloud layers over complex surfaces (e.g., cloud-snow mixtures), as expert-tuned thresholds fail to account for surface-radiation interactions that distort brightness temperature differences (BTDs)~\cite{heidinger_705_2018, marchant_evaluation_2020}. These methods struggle to account for detailed differences in complex conditions, such as mixed cloud–snow surfaces, thin overlying layers, or rapidly changing illumination, limiting their robustness under various observation conditions~\cite{zhu2012object, mahajan2020cloud}. In contrast, machine-learning (ML) methods have leveraged latent features to significantly improve multilayer cloud retrieval from satellite imagery~\cite{sun-mack_detection_2017, tan_detecting_2022, haynes_low_2022, zhao_cloud_2023}, surpassing early ANN‐based methods that merely matched threshold‐based performance~\cite{sun2017detection}. Even simple CNN-based frameworks have outperformed the operational MODIS and AHI products, achieving approximately 70\% daytime detection accuracy~\cite{li2022cloud}. However, existing CNN-based cloud retrieval models exhibit strong data dependence, due to unmodeled sensor-specific characteristics (e.g., SRF shifts)~\cite{dong2024mcdnet}

In recent years, pure data-driven deep learning methods have achieved significant progress in cloud detection. For instance, Hu et al. (2021)~\cite{hu2021cdunet} proposed CDUNet, which optimizes cloud boundaries and thin cloud localization using high-frequency feature extraction, multi-scale convolution, and a spatial prior self-attention mechanism. Ma et al. (2023)~\cite{ma2023cnn} designed CNN-TransNet with a dual-branch CNN-Transformer encoder and a differential feature enhancement module, reducing missed detections of thin clouds and false alarms over bright surfaces (e.g., snow, ice, and urban areas).
 However, these methods generally lack physical interpretability. To overcome this limitation, physics-informed deep learning methods incorporate radiative transfer physical laws or prior knowledge into the model. Heidinger et al. (2012)~\cite{heidinger2012naive} trained a naive Bayesian cloud detection scheme based on CALIPSO lidar data, establishing classifiers for different surface types and achieving good performance over both ocean and land. Li et al. (2024)~\cite{li2024physics} used clear\--sky brightness temperatures simulated by a radiative transfer model as prior information and designed a CNN\_TL model with transfer learning to simultaneously retrieve the properties of upper ice clouds and lower water clouds in multilayer clouds from Himawari\--8 passive imager observations. The retrieved cloud top heights showed strong agreement with active sensor products, outperforming the official products. These studies indicate that the cloud detection field is evolving from a pure data-driven paradigm toward a physics-data dual-driven paradigm.

To address these limitations, recent studies have resorted to physics-informed feature engineering with promising results~\cite{li_physics-driven_2024}. To further probe the interplay between physics-informed feature engineering and machine learning, three complementary lines of investigation were undertaken in this work:

\begin{enumerate}
   \item To address ambiguous physical relationships in multilayer cloud radiative transfer, our core innovation physics informed feature engineering leverages 1D-CNNs to extract high dimensional latent features from multi spectral data under physical constraints, overcoming the poor feature representation inherent in threshold-based methods.
   \item  Embedding physical priors to guide feature engineering (via threshold-derived channel selections) reduces complexity and improves interpretability—solving the trade-off between ML accuracy and operational feasibility.
  \item Translating ML-revealed physical mechanisms to threshold method optimization establishes a bidirectional synergy, overcoming the isolation of data-driven and physics-based approaches.
\end{enumerate}

\begin{figure*}[ht]
  \centering
  \includegraphics[scale=0.85]{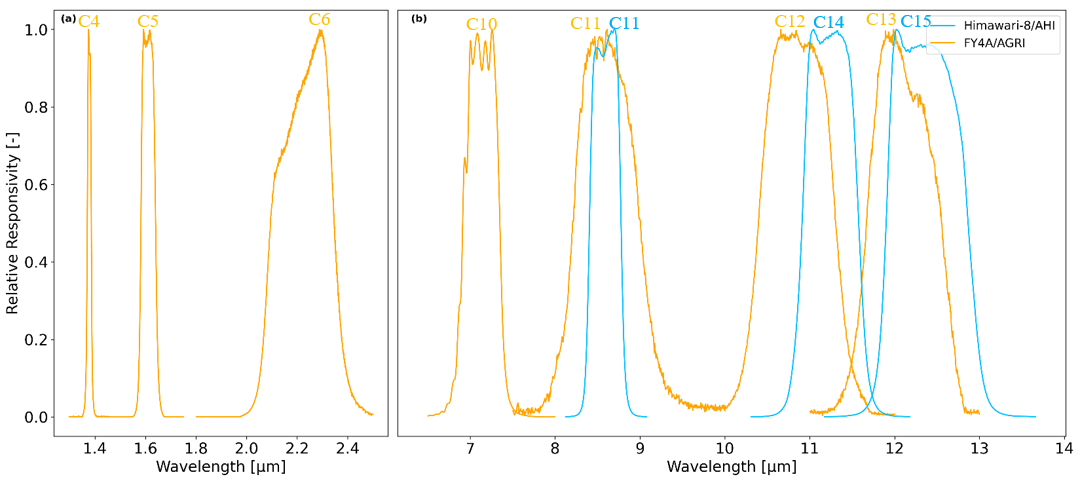}
  \caption{Spectral response functions (SRFs) of FY-4A/AGRI (orange) and Himawari-8/AHI (blue )channels. (a) Shortwave infrared (SWIR) channels. (b) Thermal infrared (TIR) channels.}
  \label{Fig1}
\end{figure*}

Section II describes the passive and active satellite observations, datasets used in this study, details of the physical methods, and the CNN-based models. Section III presents the experimental results, including quantitative performance comparisons between the 1D-CNN and threshold methods, ablation and sensitivity analyses, and validation against CloudSat–CALIOP. Section IV summarizes the main conclusions.

\section{Data and Method}
\subsection{Satellite Data}
Data for this study were obtained from Level-1 products of the Advanced Geostationary Radiation Imager (AGRI) aboard FY-4A and the Advanced Himawari Imager (AHI) aboard Himawari-8, together with the CloudSat–CALIPSO joint cloud product (CPR aboard CloudSat and CALIOP aboard CALIPSO), and are summarized in Table~\ref{tab:data}. Fig.~\ref{Fig1} indicates close agreement between AGRI and AHI spectral response functions (SRFs) near 8.5\,$\mu$m, whereas larger differences are evident around 11.0\,$\mu$m and 12.0\,$\mu$m. These sensor-specific SRF differences should be accounted for when transferring or validating channel-dependent algorithms~\cite{wang_effects_2022}.

\begin{table*}[ht]
    \centering
    \caption{Satellite datasets employed in this study. For each sensor the primary channels used, the retrieved product type (reflectance and/or brightness temperature), and the nominal spatial and temporal resolutions are listed. AGRI (FY-4A) and AHI (Himawari-8) provide multispectral imager radiances used for threshold and CNN analyses, while CALIOP (CALIPSO) and CPR (CloudSat) supply joint cloud profiles from active lidar and radar for validation and collocation.}
    \label{tab:data}
    \begin{tabular}{cccccc}
        \toprule
        Satellite & Sensor & Channels & Dataset & Spatial Resolution & Temporal Resolution  \\
        \midrule
        FY-4A &  AGRI & \makecell{C1(0.47$\mu$m);C2(0.51$\mu$m);C3(0.64$\mu$m)\\
        C4(1.375$\mu$m);C5(1.61$\mu$m);C6(2.25$\mu$m)\\ 
        C7(3.75$\mu$m);C8(3.75$\mu$m);C9(6.25$\mu$m)\\
        C10(7.1$\mu$m);C11(8.5$\mu$m);C12(10.8$\mu$m)\\
        C13(12.0$\mu$m);C14(13.5$\mu$m)
        } & Reflectance; Brightness Temperature &4km & 15min\\
        H8 &   AHI & C11(8.5$\mu$m);C14(11.2$\mu$m);C15(12.3$\mu$m) &  Reflectance; Brightness Temperature & 2km & 10min\\
        {\makecell{CALIPSO\\CloudSat }}
                  &{\makecell{CALIOP \\CPR }}
                  &{\makecell{532\&1064nm\\94GHz }}
                  &{2B-CLDCLASS-LIDAR}  & 1km      & - -                 \\
        \bottomrule
    \end{tabular}
\end{table*}

\subsubsection{Geostationary Satellites Imageries}
Observations from AGRI onboard FY-4A satellite (launched in December 2016) were employed in this study~\cite{yang_introducing_2017}. It acquires full-disk imagery every 15min via autonomous two-dimensional scanning over 14 channels~(C1–C14, ranging from 0.45$\mu$m to 13.5$\mu$m). In this study, reflectance and brightness temperatures for in October, November and December of 2018 were extracted from C4~(1.375$\mu$m) through C14~(13.5$\mu$m), covering near-infrared and longwave infrared bands at 4km resolution, along with the Cloud optical depth (COD) The data were obtained from the National Satellite Meteorological Data Center (\url{http://data.nsmc.org.cn/DataPortal/cn/home/index.html}).

The AHI aboard Himawari-8 takes full-disk imagery every 10min with channel-dependent spatial resolutions of 0.5–2km~\cite{bessho2016introduction}, including six visible/near-infrared (VNIR/SWIR) channels and ten infrared (IR) channels. For this study, full-disk Level-1 AHI products were obtained from the JAXA/EORC data portal (\url{https://www.eorc.jaxa.jp/ptree/userguide.html}) at hourly timestamps (for example, 06:00 and 07:00). The 8.5$\mu$m, 11.2$\mu$m and 12.3$\mu$m channels~\cite{zou2016characterization} were selected to compare the performance of multilayer cloud detection.

\subsubsection{CPR-CALIOP Joint Product}

Space-borne active sensors, CALIOP and CPR, provide near‐coincident observations of cloud properties. CALIOP operates at 532~nm and 1064~nm, offering high sensitivity to cloud tops, thin cirrus, and aerosol layers, but suffers strong attenuation beneath optically thick clouds~\cite{winker_calipso_2003}. CPR is a 94-GHz nadir-looking radar that penetrates upper-level clouds, making it effective for detecting multilayered formations~\cite{stephen2002cloudsat, chen_construction_2020}. However, CPR faces challenges distinguishing between aerosols and optically thin cirrus clouds. The joint cloud product of CPR and CALIOP, 2B-CLDCLASS-LIDAR, is often used to evaluate passive cloud products~\cite{oreopoulos_new_2017}. It offers a more comprehensive depiction of vertical cloud structures and reliably identifies the phase of each cloud layer~\cite{sassen2008global}. This Level 2 Radar-Lidar product is available at~\url{https://www.cloudsat.cira.colostate.edu/data-products/2b-cldclass-lidar}.

\subsection{Matched Datasets}
This study used hourly AGRI and AHI observations as input and the CPR-CALIOP joint cloud product as the reference (Fig.~\ref{Fig3}a). Temporal matching was constrained to a ±5‑minute window between the imager scan time and the active sensor overpass, balancing temporal mismatch with sufficient data retention. Spatial collocation was performed using a nearest‑neighbor matching strategy: for each imager pixel, the closest overlapping footprint from the CPR‑CALIOP product was selected. When multiple active sensor footprints fell within a single imager pixel, the lidar observations were aggregated by taking the mean for continuous variables or the mode for categorical variables to establish a one‑to‑one correspondence. When matching geostationary satellite data with CALIPSO/CloudSat, inherent uncertainties arise: CALIPSO/CloudSat measure along the vertical atmospheric column, while AGRI and AHI observe along slanted paths. As the zenith angle increases, observed feature differences become more pronounced. To minimize this uncertainty, only daytime ocean scenes from October–December 2018 with zenith angles \textless 60° were retained. This reduces biases from different observation paths, ensuring more accurate cloud retrievals. Reference labels included four classes: 0=single-layer ice, 1=multilayer, 2=single-layer water, and 3=single-layer mixed-phase. Two independent datasets were assembled: CAG, containing 130,601 CloudSat–CALIOP–AGRI pairs within the AGRI observation area, and CAA, containing 25,494 CloudSat–CALIOP–AGRI–AHI pairs within the overlapping AGRI and AHI observation area. Table~\ref{tab:results} summarizes the class distribution for both datasets. The multi-layer cloud proportions in the CAG and CAA datasets, 29\% and 36\% respectively, are consistent with typical cloud distributions in their regions and meet the experimental requirements.

\begin{table}[ht]
\centering
\caption{Matched datasets and class distribution.}
\begin{tabular}{@{}ccccc@{}} 
\toprule
Dataset Name & \makecell{Single layer\\ ice} & \makecell{Single layer\\ water} & \makecell{Single layer\\ mixed } & \makecell{Multilayer\\ cloud} \\ 
\midrule
CAG & 23K & 51K & 18K & 37K \\ 
CAA & 4K & 8K & 4K & 9K \\ 
\bottomrule
\end{tabular}
\label{tab:results}
\end{table}

\subsection{Method}
\subsubsection{Traditional Algorithm}
\begin{figure}[ht]
    \centering
    \includegraphics[width=9cm]{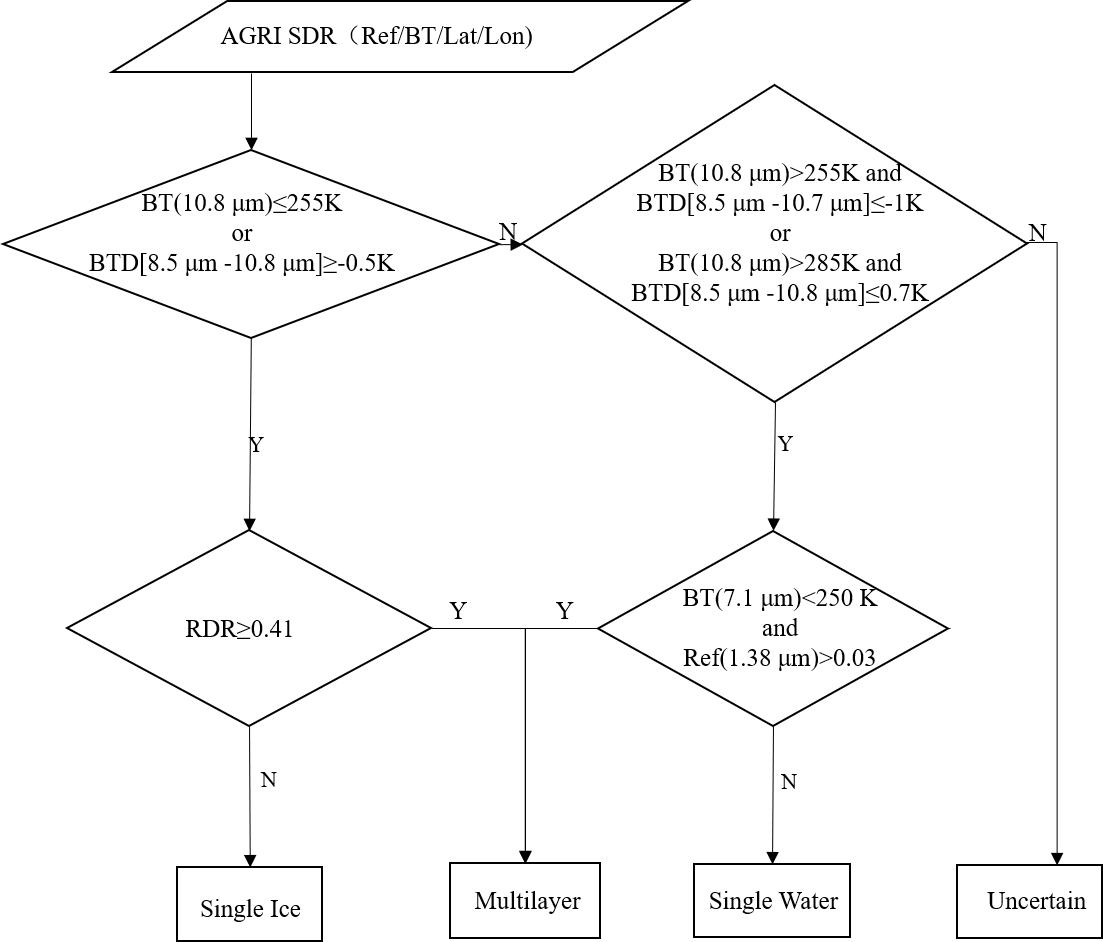}
    \caption{Flowchart of the multilayer cloud detection algorithm for FY-4A/AGRI.}
    \label{Fig2}
\end{figure}

The threshold-based framework follows the workflow described in the authors' previous study~\cite{wang_algorithm_2024} and is summarized in Fig.~\ref{Fig2}. Phase discrimination between ice and water is achieved using the brightness-temperature difference (BTD) between the 8.5$\mu$m and 10.8$\mu$m channels in combination with the 10.8$\mu$m brightness temperature (BT). The reflectance  difference ratio (RDR) between the 1.375$\mu$m and 1.61$\mu$m channels, defined in Eq.~(1), is employed to identify multilayer scenes in which an ice cloud overlies a water cloud. When the top layer is classified as water, a joint discriminator based on the 7.1$\mu$m BT and the 1.375$\mu$m reflectance is applied to separate single-layer water clouds from multilayer configurations.

\begin{equation}
RDR = \frac{R_{1.61\,\mu\text{m}} - R_{1.375\,\mu\text{m}}}{R_{1.375\,\mu\text{m}} + R_{1.61\,\mu\text{m}}}
\end{equation}

\subsubsection{CNN-based Algorithm}
\begin{figure*}[htp] 
\centering 
\includegraphics[scale=0.42]{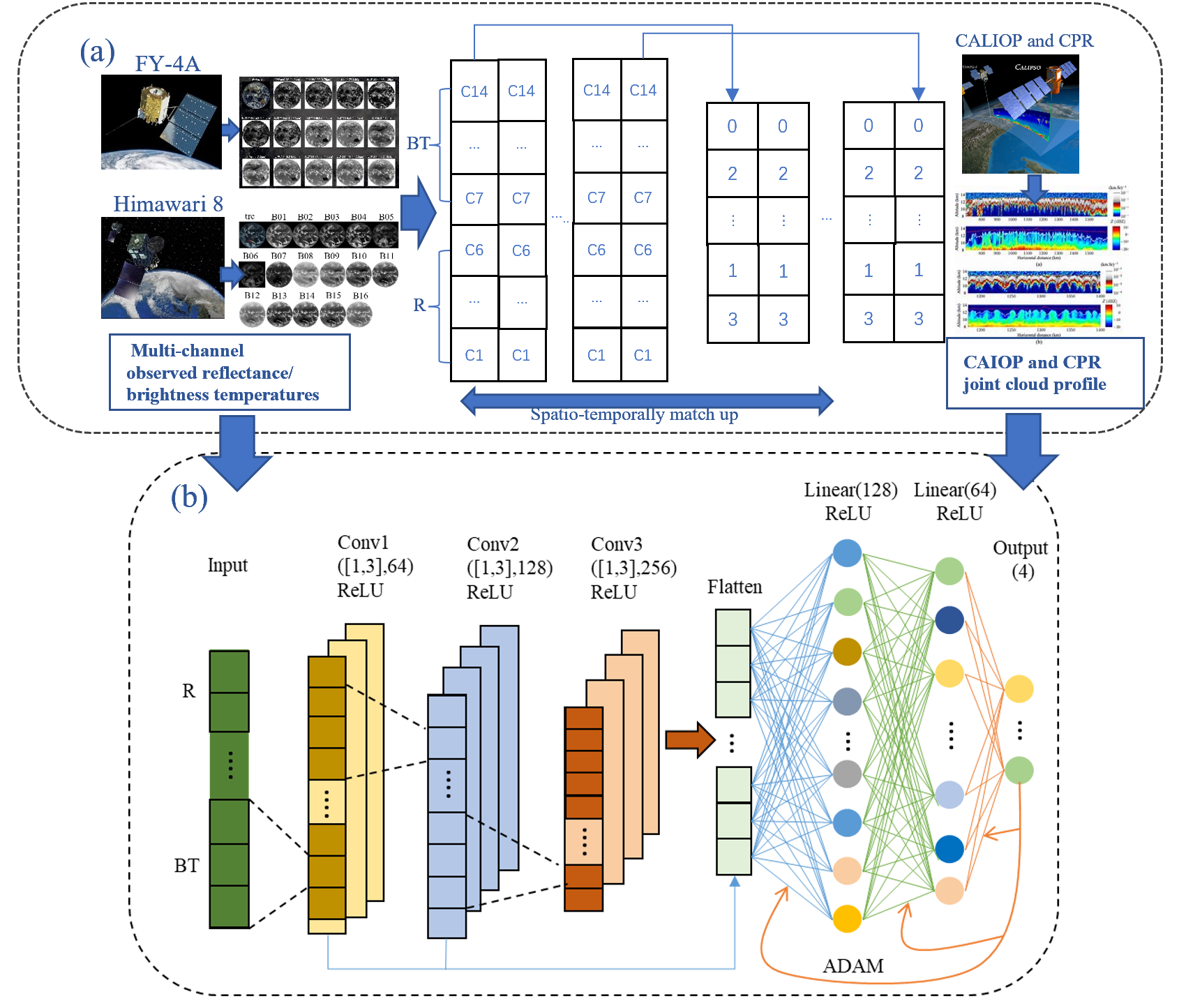} 
\caption{(a) The flowchart for generating the dataset. (b) Model configuration of the deep neural network (CNN).} 
\label{Fig3} 
\end{figure*}

Motivated by prior CNN-based approaches to cloud retrieval~\cite{tan_detecting_2022, li2022cloud}, a one-dimensional convolutional network was designed. The architecture comprises an input layer, three 1D convolutional layers, two fully connected layers, and a final softmax output layer that yields class probabilities (see Fig.~\ref{Fig3}(b)). The number of convolutional stages was selected via ablation experiments to balance representational power and model compactness. Each convolutional layer employs a kernel size of 3, which after three successive convolutions yields a receptive field sufficient to capture the relevant spectral correlations while keeping parameter growth minimal. Batch normalization is applied after each convolution to stabilize and accelerate training, ReLU activations introduce nonlinearity and mitigate vanishing gradients, and a categorical cross-entropy loss is minimized using the Adam optimizer. This configuration was chosen to provide an efficient, interpretable backbone suitable for operational deployment. The proposed network architecture contains approximately 116k trainable parameters. On the AGRI L1 dataset, it achieves an inference time of about 242 seconds and a memory usage of roughly 10.4 MB.

\subsubsection{Experimental Design and Evaluation}

Experiments on the CAG dataset employed a baseline model, denoted CNN-PhysCore, constructed from the same channel set as the Baseline Algorithm, or the threshold-based algorithm of AGRI, 
to enable a direct physics–ML comparison. Two partial-convolution experiments, namely CNN-CoreConv and CNN-AuxConv were implemented to investigate feature routing. In CNN-CoreConv, the core channels (identical to those used in CNN-PhysCore) are processed through the convolutional layers, while auxiliary channels are supplied to the fully connected branch. By contrast, in CNN-AuxConv the auxiliary channels are routed through the convolutional layers, and the core channels are directly linked to the fully connected layer.Channel-selection effects were evaluated by comparing CNN-Gradient, whose input was restricted to the five channels with the highest gradient-based importance scores, against CNN-PhysCore. Building on these insights, a further comparison was conducted between CNN-PhysCore-C13Add (the PhysCore set with AGRI C13 at 12.0\,$\mu$m appended) and CNN-PhysCore-C13Sub (with C12 at 10.8\,$\mu$m replaced by C13). In addition, a nighttime model (CNN-night) was designed using only water vapor and thermal infrared channels. All experimental details are listed in Table~\ref{tab 4} and Table~\ref{tab 5}.
Specifically, we provide a multilayer cloud detection algorithm for nighttime scenes, denoted CNN\--night, along with detailed evaluation results over different underlying surfaces for the CNN\--SWIR2TIR configuration, as shown in Fig. S1.

For benchmarking, a set of threshold experiments was defined for cross-sensor comparison. Baseline-AGRI denotes the original AGRI threshold algorithm, while AGRI-SubC13 denotes the AGRI baseline algorithm with C12 replaced by C13. AHI-SubC14 denotes the AHI threshold algorithm obtained by mapping AGRI's thermal-infrared tests to the most similar AHI channels ( AGRI 8.5\,\(\mu\)m \(\rightarrow\) AHI 8.5\,\(\mu\)m, AGRI 10.8\,\(\mu\)m \(\rightarrow\) AHI 11.2\,\(\mu\)m). AHI-SubC15 is similar to the AHI-SubC14 experiment but with channel C15( AGRI 8.5\,\(\mu\)m \(\rightarrow\) AHI 8.5\,\(\mu\)m, AGRI 12.0\,\(\mu\)m \(\rightarrow\) AHI 12.3\,\(\mu\)m). Full experimental configurations are summarized in Table~\ref{tab 6}.

The evaluation metrics comprise the probability of detection (POD) and false alarm rate (FAR) for multilayer and single-layer clouds, together with two aggregate scores: hit rate (HR) and the Hanssen–Kuipers skill score (KSS). 

Two phase-conditioned recognition ratios were computed to clarify the two-stage behaviour of the threshold algorithms. MwI denotes the fraction of multilayer scenes correctly identified when the top layer is classified as ice, and MwW denotes the fraction correctly identified when the top layer is classified as water. These ratios were used to attribute performance to specific channels by separating (a) the first-stage phase discrimination (ice vs. water) from (b) the subsequent phase-conditioned multilayer test, thereby revealing which channels primarily support top-layer identification and which drive detection of underlying or overlying layers.
The metrics are given by
\begin{align}
\mathrm{POD}_{\rm mul} &= \frac{a}{a+b}, \\
\mathrm{FAR}_{\rm mul} &= \frac{c}{a+c}, \\
\mathrm{POD}_{\rm sin} &= \frac{d}{c+d}, \\
\mathrm{FAR}_{\rm sin} &= \frac{b}{b+d}, \\
\mathrm{HR} &= \frac{a+d}{a+b+c+d},  \\
\mathrm{KSS} &= \frac{a}{a+b}-\frac{c}{c+d} \;=\; \frac{ad-bc}{(a+b)(c+d)};\\
\mathrm{MwI} & = \frac{m}{a+b};\\
\mathrm{MwW} & = \frac{n}{a+b}.
\end{align}
 In Table~\ref{tab 3}
\(a\) means true multilayer (by both algorithm and reference), 
\(b\) means reference multilayer missed by the algorithm, 
\(c\) means false multilayer, single-layer but marked as multilayer by algorithm, 
\(d\) means true single-layer. 
The phase-conditioned subsets \(m\) and \(n\) denote the portions of \(a\) for which the algorithm’s top layer is ice and water, respectively (\(a=m+n\)).

\begin{table}[!htbp]
\centering
\caption{Simplified confusion matrix for multilayer cloud detection.}
\footnotesize  
\begin{tabular}{lcc} 
\hline 
\diagbox{Prediction}{Reference} & \shortstack{Multi- layered} & \shortstack{Single-layer} \\ 
\hline
Multilayered & m+n=a & c \\
\hline
Single-layer & b & d \\
\hline 
\end{tabular}
\label{tab 3}
\end{table}
POD (recall) ranges from 0 to 1 (higher is better); FAR is ideally near 0. HR denotes overall accuracy and KSS is a bias-adjusted skill score (range \(-1\) to \(1\)). Further evaluation details are given in Section~III.

\section{Results and Discussion}
\subsection{CNN Architectures and Channel Selection}
 
\begin{table*}[ht]
    \centering
    \caption{Performance comparison of baseline and CNN models with varied convolution and parameter configurations.} 
    \begin{tabular}{ccccccc}
        \toprule
        models &input& POD\textsubscript{mul} & FAR\textsubscript{mul}  & HR &KSS & parameter  \\
        \midrule
       Baseline-AGRI& $R_{1.375\,\mu\mathrm{m}}$,$R_{1.61\,\mu\mathrm{m}}$,$BT_{7.1\,\mu\mathrm{m}}$,$BT_{8.5\,\mu\mathrm{m}}$,$BT_{10.8\,\mu\mathrm{m}}$ & 0.558& 0.369 &0.688 &0.350 & -  -  -\\
        CNN-PhysCore & Conv($R_{1.375\,\mu\mathrm{m}}$,$R_{1.61\,\mu\mathrm{m}}$,$BT_{7.1\,\mu\mathrm{m}}$,$BT_{8.5\,\mu\mathrm{m}}$,$BT_{10.8\,\mu\mathrm{m}}$) &  0.549 & 0.301 & 0.744 & 0.408 & 115716 \\
        CNN-SWIR2TIR & \makecell{Conv($R_{1.375\,\mu\mathrm{m}}$,$R_{1.61\,\mu\mathrm{m}}$,$R_{2.25\,\mu\mathrm{m}}$,$BT_{3.75\,\mu\mathrm{m}}$,$BT_{6.1\,\mu\mathrm{m}}$\\$BT_{7.1\,\mu\mathrm{m}}$,$BT_{8.5\,\mu\mathrm{m}}$,$BT_{10.8\,\mu\mathrm{m}}$,$BT_{12.0\,\mu\mathrm{m}}$,$BT_{13.5\,\mu\mathrm{m}}$} & 0.620 & 0.251 & 0.781 & 0.497 & 115716 \\
        CNN-CoreConv & \makecell{Conv($R_{1.375\,\mu\mathrm{m}}$,$R_{1.61\,\mu\mathrm{m}}$,$BT_{7.1\,\mu\mathrm{m}}$,$BT_{8.5\,\mu\mathrm{m}}$,$BT_{10.8\,\mu\mathrm{m}}$) \\ $R_{2.25\,\mu\mathrm{m}}$,$BT_{3.75\,\mu\mathrm{m}}$,$BT_{6.9\,\mu\mathrm{m}}$,$BT_{12.0\,\mu\mathrm{m}}$,$BT_{13.5\,\mu\mathrm{m}}$} & 0.600 & 0.278 & 0.765 & 0.465 & 118020 \\
        CNN-AuxConv & \makecell{Conv($R_{2.25\,\mu\mathrm{m}}$,$BT_{3.75\,\mu\mathrm{m}}$,$BT_{6.9\,\mu\mathrm{m}}$,$BT_{12.0\,\mu\mathrm{m}}$,$BT_{13.5\,\mu\mathrm{m}}$)\\+$R_{1.375\,\mu\mathrm{m}}$,$R_{1.61\,\mu\mathrm{m}}$,$BT_{7.1\,\mu\mathrm{m}}$,$BT_{8.5\,\mu\mathrm{m}}$,$BT_{10.8\,\mu\mathrm{m}}$} & 0.605 & 0.241 & 0.781 & 0.453 & 116996  \\
     
        \bottomrule
    \end{tabular}
    \label{tab 4}
\end{table*}

\begin{figure}[ht]
    \centering
    \includegraphics[width=8cm]{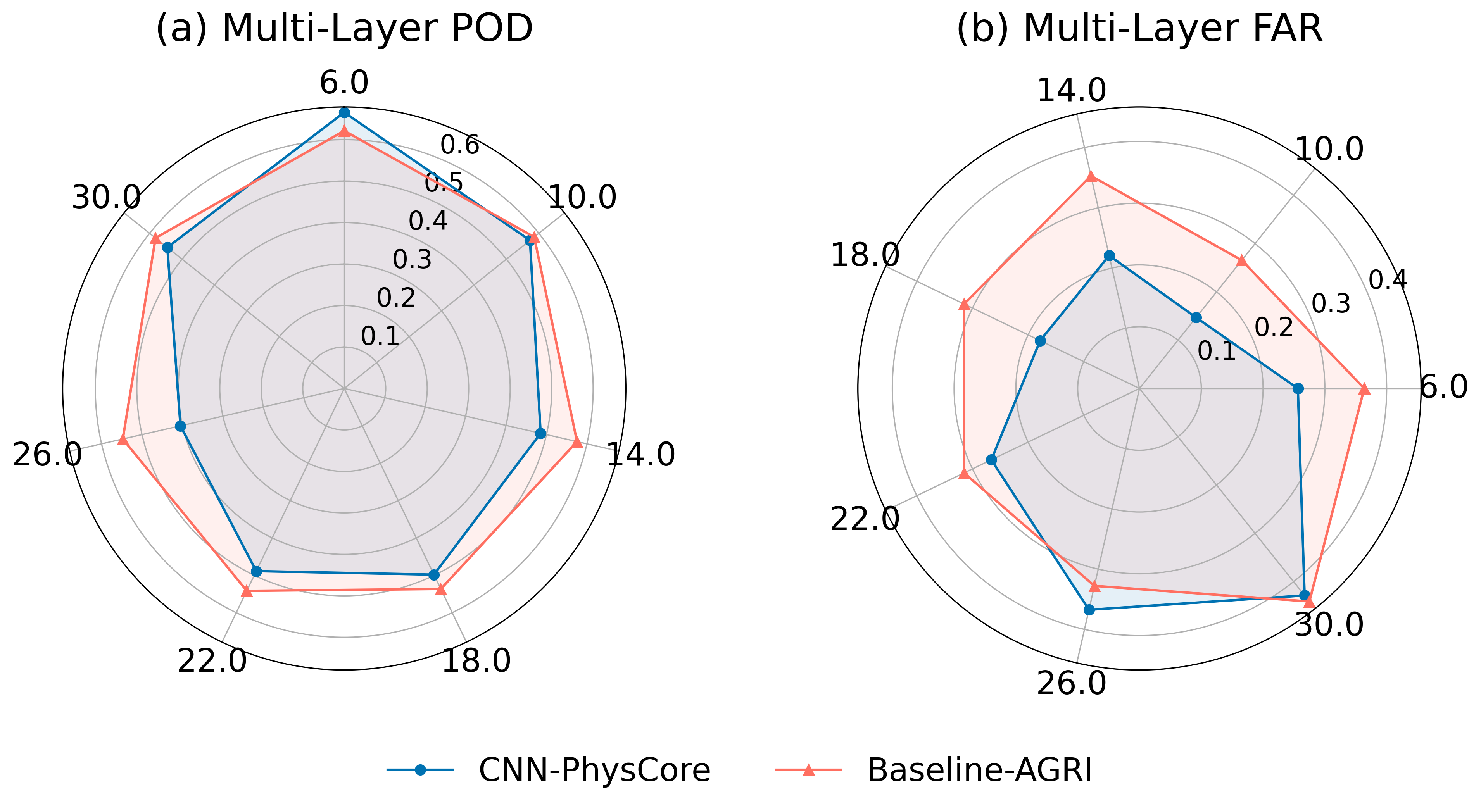}
    \caption{Variation of probability of detection (POD) and false alarm rate (FAR) with cloud optical thickness(COD) for CNN-PhysCore and the Baseline algorithm(threshold). (a) POD of multilayer cloud detection. (b) FAR of multilayer cloud detection.}
     
    \label{Fig 4}
\end{figure}

The results demonstrate that the ML model (CNN-PhysCore) and the threshold baseline achieve comparable multilayer detection rates (POD$_{\rm mul}$) across most optical-thickness regimes, as illustrated in Fig.~\ref{Fig 4}. However, CNN-PhysCore consistently yields a lower false-alarm rate (FAR$_{\rm mul}$), indicating better discriminative ability. Consistent with Table~\ref{tab 4}, compared with the Baseline algorithm (POD$_{\rm mul}=0.558$, FAR$_{\rm mul}=0.369$, HR$=0.688$), CNN-PhysCore achieves a similar POD$_{\rm mul}=0.549$ while reducing FAR$_{\rm mul}$ to $0.301$ and improving HR to $0.744$.
When supplied with the full SWIR-to-TIR channel suite (1.375--13.5\,$\mu$m), the CNN-SWIR2TTR model produced the best detection results among all evaluated models, as summarized in Table~IV. It attains POD$_{\rm mul}=0.620$ and FAR$_{\rm mul}=0.251$ (parameters $\approx115{,}716$). However, increasing the number of input channels introduces not only potentially complementary information but also possible noise and redundancy. To systematically investigate how the flow of information through the convolutional pathway affects detection performance, we designed two partial-convolution experiments: CNN-CoreConv and CNN-AuxConv. In CNN-CoreConv (parameters $\approx118{,}020$), the core channels (identical to those used in CNN-PhysCore) are processed by convolutional layers, while the auxiliary channels bypass convolutions and are fed directly to the fully connected branch, achieving POD$_{\rm mul}=0.600$ and FAR$_{\rm mul}=0.278$. In contrast, CNN-AuxConv (parameters $\approx116{,}996$) sends the auxiliary channels through the convolutional pathway and routes the core channels directly to the fully connected layer, attaining POD$_{\rm mul}=0.605$ and FAR$_{\rm mul}=0.241$. The superior performance of CNN-AuxConv suggests that the core channels already possess highly discriminative features directly useful for classification; applying additional convolution to them could instead introduce noise and degrade accuracy. Meanwhile, the auxiliary channels benefit from convolutional processing to extract higher-dimensional cues. Hence, routing core channels directly to the fully connected layer preserves their inherent signal, while feeding auxiliary channels into the convolutional pathway enables effective feature extraction.

\begin{figure*}[htp]
    \centering
    \includegraphics[width=\textwidth]{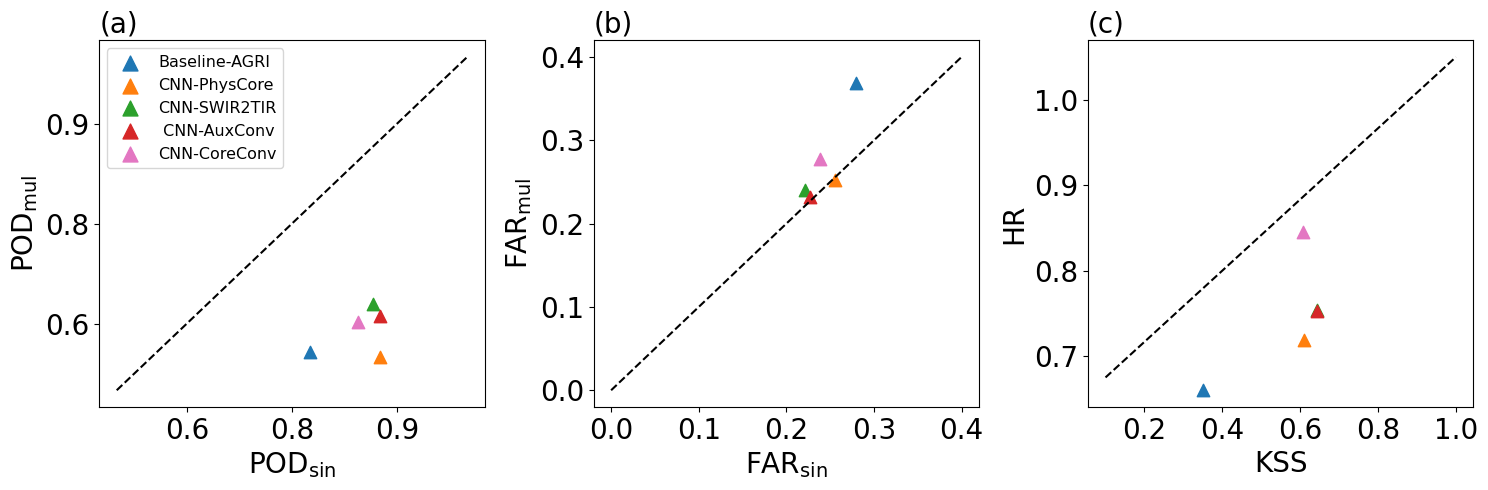}
    \caption{Scatter comparison of evaluation metrics for multilayer versus single-layer cloud detection algorithms based on the 2018 CPR–CALIOP merged cloud product. (a) Probability of detection (POD\textsubscript{multi} vs. POD\textsubscript{single}). (b) False alarm rate (FAR\textsubscript{multi} vs. FAR\textsubscript{single}). (c) Hit rate (HR) versus Hanssen–Kuipers skill score (KSS).}
    \label{Fig 5}
\end{figure*}

Fig.~\ref{Fig 5} complements Table~\ref{tab 4} by reporting POD$_{\rm mul}$, POD$_{\rm sin}$, FAR$_{\rm mul}$, FAR$_{\rm sin}$, HR, and KSS for each experiment. The plots confirm that CNN-SWIR2TIR attains the highest overall skill (KSS $=0.497$) and that CNN-AuxConv closely matches this performance (KSS $=0.453$). The full-channel model CNN-SWIR2TIR applies convolution to all input bands, including the core channels. This operation improves multilayer detection accuracy to a certain extent but simultaneously introduces noise, resulting in a slightly higher FAR$_{\rm mul}$ ($0.251$) compared to CNN-AuxConv ($0.241$). Single-layer metrics are consistent with the multilayer results, indicating the overall robustness of the experiments. Notably, FAR$_{\rm mul}$ systematically exceeds FAR$_{\rm sin}$ across experiments, reflecting the greater intrinsic difficulty of multilayer detection. This reflects the complex radiative interactions in multilayer scenes: the upper layer frequently masks lower-layer signals, producing ambiguous TOA radiances and thereby increasing FAR for passive retrievals~\cite{sun-mack_identification_2024}.

\subsection{Physical method and data-driven collaborative optimization} 

\begin{figure}[ht]
    \centering
    \includegraphics[width=8cm]{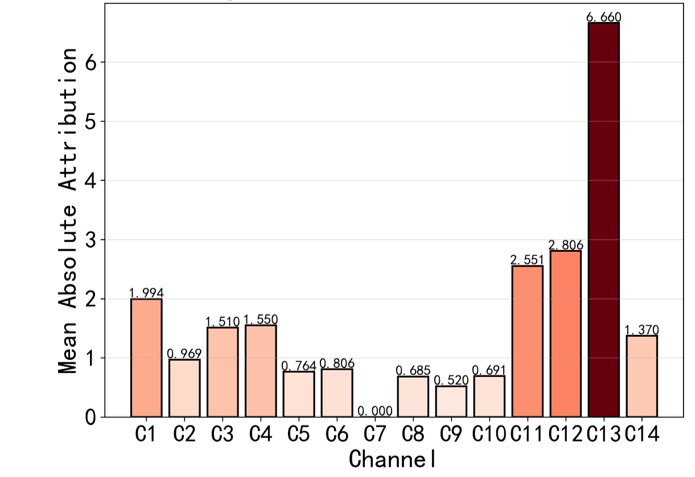}
    \caption{Feature gradient (mean absolute attribution) for all channels in the CNN-SWIR2TIR experiment.}
     
    \label{Fig 6}
\end{figure}

A feature-gradient analysis (Fig.~\ref{Fig 6}) identified AGRI C13 (12.0\,$\mu$m) as the most influential channel for multilayer detection, followed by C1 (0.47\,$\mu$m), C12 (10.8\,$\mu$m), C11 (8.5\,$\mu$m), and C4 (1.375\,$\mu$m). Guided by these rankings, a gradient-only configuration (CNN-Gradient) using these five channels, together with two experiments (CNN-PhysCore-C13Add and CNN-PhysCore-C13Sub), were evaluated (Table~\ref{tab 5}) to test whether ML-identified importance yields tangible gains. In addition, a separate configuration designed for nighttime scenes, consisting of C13, C12, C11, C10 (7.1\,$\mu$m), and C9 (6.1\,$\mu$m), was also examined (referred to as the nighttime model); this set is independent of the gradient-based ranking.

\begin{table*}[ht]
    \centering
    \caption{Performance metrics and parameter counts for CNN models with varied parameter configurations.} 
    \begin{tabular}{lllllll}
        \toprule
        models &input& POD\textsubscript{mul} & FAR\textsubscript{mul}  & HR &KSS & parameter  \\
        \midrule
        CNN-Gradient-Feat & \makecell{Conv($R_{0.47\,\mu\mathrm{m}}$,$R_{1.375\,\mu\mathrm{m}}$,$BT_{8.5\,\mu\mathrm{m}}$\\$BT_{10.8\,\mu\mathrm{m}}$,$BT_{12.0\,\mu\mathrm{m}}$)} & 0.620 & 0.240 & 0.776 & 0.386 & 115716  \\
        CNN-PhysCore-C13Add & \makecell{Conv($R_{1.375\,\mu\mathrm{m}}$,$R_{1.61\,\mu\mathrm{m}}$,$BT_{7.1\,\mu\mathrm{m}}$\\$BT_{8.5\,\mu\mathrm{m}}$,$BT_{10.8\,\mu\mathrm{m}}$,$BT_{12.0\,\mu\mathrm{m}}$)} & 0.537 & 0.279 & 0.750 & 0.413 & 115716  \\
        CNN-PhysCore-C13Sub & \makecell{Conv($R_{1.375\,\mu\mathrm{m}}$,$R_{1.61\,\mu\mathrm{m}}$,$BT_{7.1\,\mu\mathrm{m}}$\\$BT_{8.5\,\mu\mathrm{m}}$,$BT_{12.0\,\mu\mathrm{m}}$)} & 0.562 & 0.287 & 0.753 & 0.428 & 115716  \\   
        CNN-night & \makecell{Conv($BT_{6.25\,\mu\mathrm{m}}$,$BT_{7.1\,\mu\mathrm{m}}$,$BT_{8.5\,\mu\mathrm{m}}$\\$BT_{10.8\,\mu\mathrm{m}}$,$BT_{12.0\,\mu\mathrm{m}}$)} & 0.456 & 0.276 & 0.733 & 0.352 & 115716  \\
       
        \bottomrule
    \end{tabular}
    \label{tab 5}
\end{table*}

Compared with Baseline-AGRI, the data-driven gradient experiment (CNN-Gradient-Feat) substantially improves all metrics (POD$_{\rm mul}=0.620$, FAR$_{\rm mul}=0.240$, HR$=0.776$, KSS$=0.386$) and outperforms CNN-PhysCore. Its channel configuration largely overlaps with that of CNN-PhysCore, indicating consistency between prior physical knowledge and data-driven attribution. It achieves performance comparable to the full-channel CNN-SWIR2TIR (POD$_{\rm mul}=0.620$, FAR$_{\rm mul}=0.251$) while reducing the multilayer false-alarm rate to $0.240$. This advantage stems from gradient-based channel selection, which retains informative bands and reduces noisy inputs. In contrast, the CNN-night model, which excludes near-infrared bands, yields the poorest performance (POD$_{\rm mul}=0.456$, FAR$_{\rm mul}=0.276$), consistent with the physical prior that multilayer cloud detection inherently depends on near-infrared channels.

Appending AGRI C13 (CNN-PhysCore-C13Add) yields POD$_{\rm mul}=0.537$, FAR$_{\rm mul}=0.279$, HR$=0.750$, KSS$=0.413$ (parameters $\approx115{,}716$), while replacing C12 with C13 (CNN-PhysCore-C13Sub) achieves POD$_{\rm mul}=0.562$, FAR$_{\rm mul}=0.287$, HR$=0.753$, KSS$=0.428$ (parameters $\approx115{,}716$). The improvement of CNN-PhysCore-C13Sub over CNN-PhysCore-C13Add suggests that substituting C12 with C13 is more effective than simply appending C13, likely because both C12 and C13 sample thermal-infrared window physics and respond comparably to cloud emissivity and TOA radiance differences~\cite{pavolonis2004daytime,heidinger_705_2018}. Additionally, C13 exhibits sensitivity to cloud top height and ice-particle emissivity, aiding multilayer cloud detection.

\begin{figure*}[htp]
    \centering
    \includegraphics[scale=0.5]{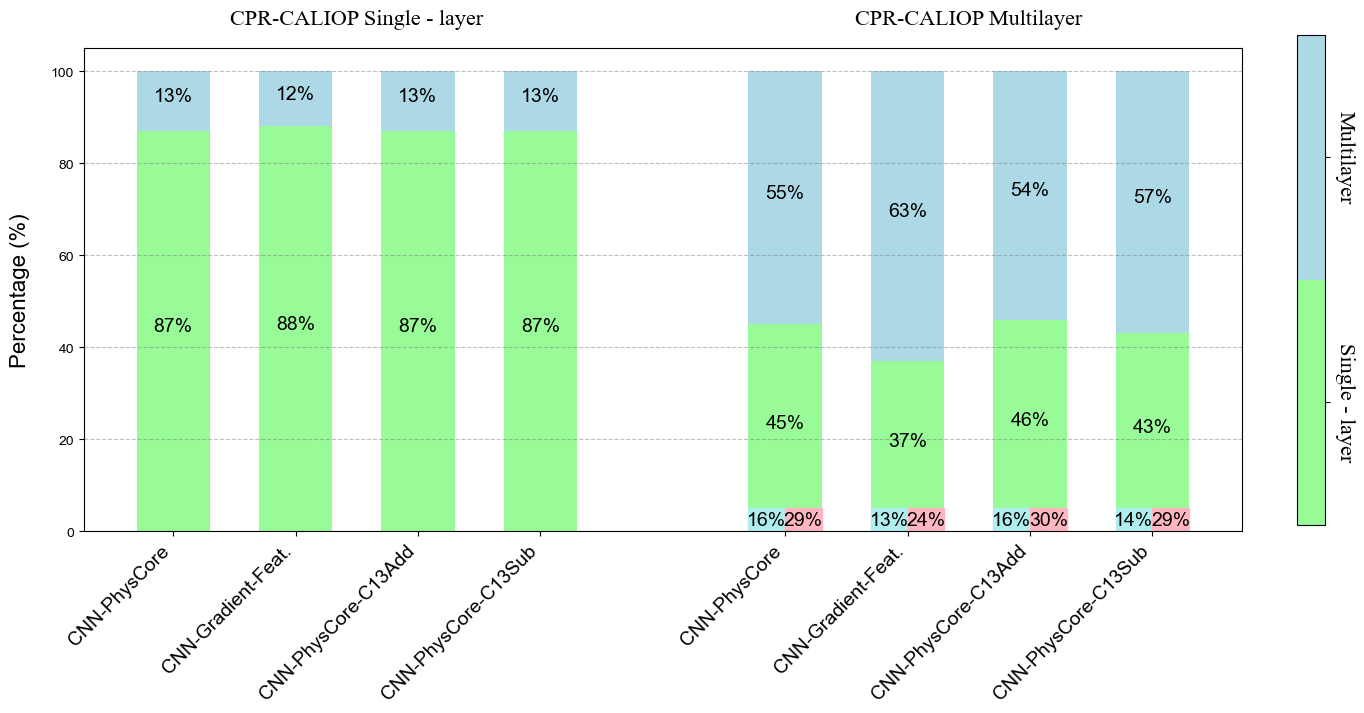}
    \caption{The percentages of single-layer and multilayer clouds for products of 1D-CNN models with different parameters 
     of the CPR-CALIOP reference labels. The figure also lists the percentage of multilayer clouds misclassified as single-layer ice (light cyan), mixed-phase (beige), or water clouds (pale pink) by different experiments.
     }
    \label{Fig 7}
\end{figure*}

Validation was restricted to samples with total cloud optical thickness (COT) \(>5\) and employed the CPR–CALIPSO joint product as the reference (see Fig.~\ref{Fig 7}). Under single-layer conditions, the data-driven experiment CNN-Gradient-Feat attains the highest single-layer detection accuracy among all models (\(\approx88\%\)), outperforming the other three configurations. For multilayer conditions, CNN-Gradient-Feat also achieves the best performance, with a multilayer detection accuracy of \(\approx63\%\). Its principal misclassifications are single-layer ice clouds (\(\approx13\%\)) and single-layer water clouds (\(\approx24\%\)), with the latter being significantly more frequent. These results demonstrate that data-driven attribution analysis can extract informative features for classification while reducing noisy channel inputs, thereby improving overall detection performance.

By contrast, the hybrid experiment CNN-PhysCore-C13Add yields performance comparable to the baseline CNN-PhysCore in both single-layer and multilayer detection, suggesting that simply appending C13 does not bring a clear benefit. In comparison, CNN-PhysCore-C13Sub achieves a higher multilayer detection accuracy (57\%) than CNN-PhysCore (55\%), primarily by reducing misclassifications as single-layer ice clouds from 16\% to 14\%. This improvement indicates that for a fixed parameter configuration, adding a channel without prior physical insight may introduce noise, limiting the effectiveness of data-driven learning. Drawing on prior physical knowledge of the similarity between C12 and C13, the substitution strategy proves more effective, yielding a superior configuration compared to the original CNN-PhysCore.
Appending AGRI C13 (12.0\,$\mu$m) to the PhysCore set does not yield a clear improvement, whereas substituting C12 (10.8\,$\mu$m) with C13 consistently enhances multilayer detection performance. This result aligns with the ML-identified importance of C13 and is physically plausible: both C12 and C13 sample the thermal-infrared window, but the 12.0\,$\mu$m channel is more sensitive to ice-crystal emissivity and particle-size–dependent spectral variations, whereas the 10.8\,$\mu$m channel primarily reflects surface/cloud-top temperature. Their joint use has been demonstrated to improve discrimination between overlying ice and underlying water clouds~\cite{pavolonis2004daytime}. The substitution strategy, grounded in prior physical knowledge of channel similarity, proves more effective than simply appending the new channel.

\subsection{Impact of Channel Stability on Algorithm Generalization}
Previous results indicated the potential utility of replacing AGRI C12 with C13 in threshold tests. Building on that insight and on prior work highlighting substantial on-orbit radiometric variability of AGRI thermal channels~\cite{wang_effects_2022}, the baseline threshold method was therefore adapted to exploit AGRI C13 within the two-stage workflow of the baseline algorithm (top-layer phase identification followed by a phase-conditioned check for underlying/overlying clouds). To assess transferability and sensor-specific effects, analogous substitutions were applied to Himawari-8 ( 'SubC14' and 'SubC15' experiments) based on inter-sensor SRF comparisons. Phase-conditioned performance was summarized using two indices: MwI (correct multilayer detections when the top layer is identified as ice) and MwW (the same when the top layer is identified as water) (detailed in Eqs.~(8)and Eqs.~(9) respectively). Evaluations explicitly account for sensor SRF differences and on-orbit stability, which can limit the effectiveness of simple channel substitutions.

\begin{equation}
MwI = \frac{m}{a+b};
\end{equation}
\begin{equation}
MwW = \frac{n}{a+b}.
\end{equation}

\begin{figure*}[htp]
    \centering
    \includegraphics[width=\textwidth]{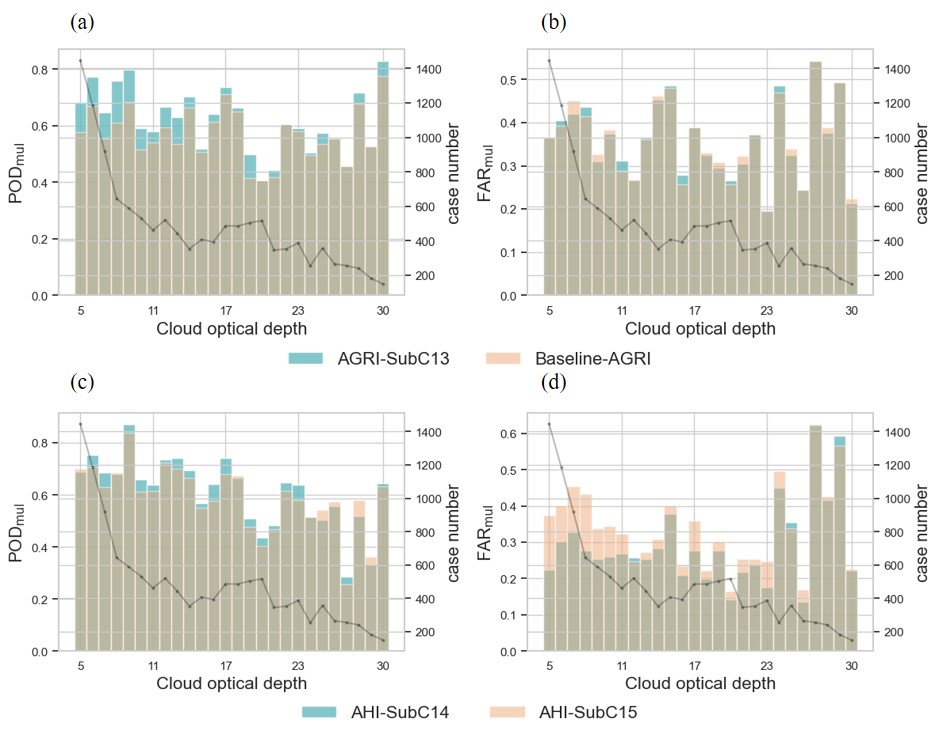}
    \caption{Comparison of POD and FAR for threshold-based algorithms against the CPR-CALIOP merged cloud product. (a,b) POD and FAR for the baseline algorithm and the SubC13 threshold algorithm. (c,d) POD and FAR for SubC14 and SubC15 threshold algorithms. Dashed lines show the number of sample points in each cloud optical depth bin (CAA dataset, Section 2).}
    \label{Fig 8}
\end{figure*}

Table~\ref{tab 6} summarizes the performance of four threshold experiments ( Baseline-AGRI, AGRI-SubC13, AHI-SubC14, AHI-SubC15) using POD$_{\mathrm{mul}}$, FAR$_{\mathrm{mul}}$, HR, KSS, and the phase-conditioned recognition ratios MwI (multilayer correctly identified when the top layer is ice) and MwW (multilayer correctly identified when the top layer is water).  
\begin{table*}[ht]
    \centering
    \caption{Performance metrics for threshold-based multilayer cloud detection algorithms with different channel configurations.} 
    \begin{tabular}{lllllllll}
        \toprule
        models &input& POD\textsubscript{mul} & FAR\textsubscript{mul}  & HR &KSS &MwI&MwW \\
        \midrule
        Baseline-AGRI & \makecell{$R_{1.375\,\mu\mathrm{m}}$,$R_{1.61\,\mu\mathrm{m}}$\\$BT_{7.1\,\mu\mathrm{m}}$,$BT_{8.5\,\mu\mathrm{m}}$,$BT_{10.8\,\mu\mathrm{m}}$} &  0.558 & 0.369 & 0.688 & 0.350 & 0.352&0.206\\
        AGRI-SubC13  & \makecell{$R_{1.375\,\mu\mathrm{m}}$,$R_{1.61\,\mu\mathrm{m}}$\\$BT_{7.1\,\mu\mathrm{m}}$,$BT_{8.5\,\mu\mathrm{m}}$,$BT_{12.0\,\mu\mathrm{m}}$} &  0.609 & 0.368 & 0.697 & 0.371 & 0.476&0.133 \\
        AHI-SubC15  & \makecell{$R_{1.375\,\mu\mathrm{m}}$,$R_{1.61\,\mu\mathrm{m}}$\\$BT_{7.1\,\mu\mathrm{m}}$,$BT_{8.5\,\mu\mathrm{m},\,\mathrm{AHI}}$,$BT_{11.2\,\mu\mathrm{m},\,\mathrm{AHI}}$} &  0.600 & 0.349 & 0.707 & 0.391 & 0.564&0.036 \\
        AHI-SubC14  & \makecell{$R_{1.375\,\mu\mathrm{m}}$,$R_{1.61\,\mu\mathrm{m}}$\\$BT_{7.1\,\mu\mathrm{m}}$,$BT_{8.5\,\mu\mathrm{m},\,\mathrm{AHI}}$,$BT_{12.3\,\mu\mathrm{m},\,\mathrm{AHI}}$} &  0.621 & 0.286 & 0.745 & 0.475 & 0.567&0.054 \\
        \bottomrule
    \end{tabular}
    \label{tab 6}
\end{table*}

Since POD$_{\mathrm{mul}}$ is jointly determined by MwI and MwW, multilayer detection is found to depend primarily on MwI because near-infrared diagnostics are effective at revealing water clouds beneath ice. Comparison shows that SubC14’s higher POD$_{\mathrm{mul}}$ is chiefly driven by an increased MwI, whereas the Baseline attains a larger MwW but at the cost of a higher FAR$_{\mathrm{mul}}$. Introducing AGRI C13 strengthens top-layer ice identification, improving the phase split and multilayer retrieval; applying the same substitution to AHI yields sensor-dependent differences attributable to SRF and on-orbit radiometric characteristics rather than to algorithmic design.

\begin{figure*}[htp]
    \centering
    \includegraphics[width=\textwidth]{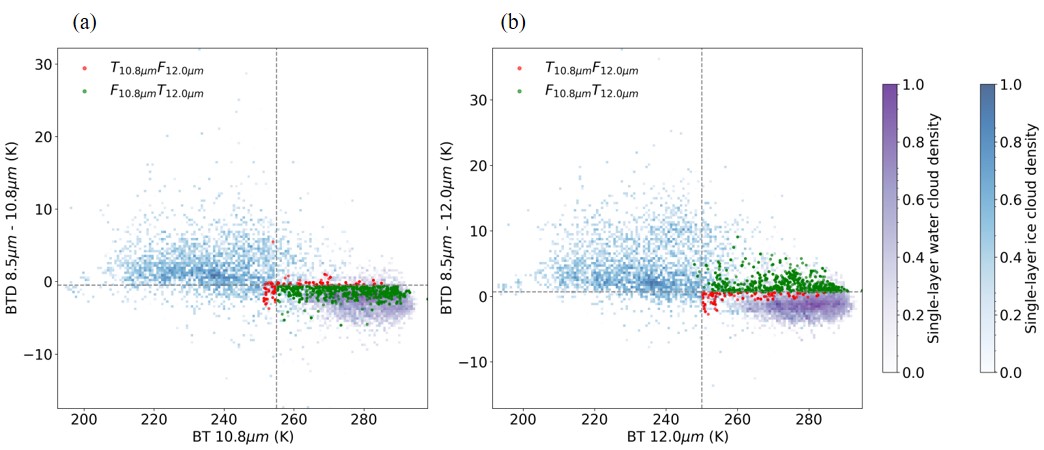}
    \caption{Joint distributions of brightness temperature (BT) and brightness-temperature difference (BTD) used by the two-stage threshold scheme.  Each panel displays per-pixel scatter samples colored by reference class (single-layer ice, single-layer water) and annotated decision thresholds (dashed lines).  (a) Baseline algorithm (AGRI): horizontal axis — BT at 10.8\,$\mu$m; vertical axis — BTD$_{8.5-10.8}$ (BT$_{8.5\mu\mathrm{m}}-$BT$_{10.8\mu\mathrm{m}}$). The baseline decision thresholds are indicated by the dashed lines (BT$_{10.8}=255$\,K and BTD$_{8.5-10.8}=-0.5$\,K).  (b) AGRI-SubC13 variant: horizontal axis — BT at 12.0\,$\mu$m; vertical axis — BTD$_{8.5-12.0}$ (BT$_{8.5\mu\mathrm{m}}-$BT$_{12.0\mu\mathrm{m}}$). The SubC13 decision thresholds are shown by dashed lines (BT$_{12.0}=250$\,K and BTD$_{8.5-12.0}=0.7$\,K). 
    Green markers denote multilayer‐cloud samples detected only by the 12.0\,$\mu$m threshold; red markers denote those detected only by the 10.8\,$\mu$m threshold.}
    \label{Fig 9}
\end{figure*}

\begin{figure*}[htp]
    \centering
    \includegraphics[width=\textwidth]{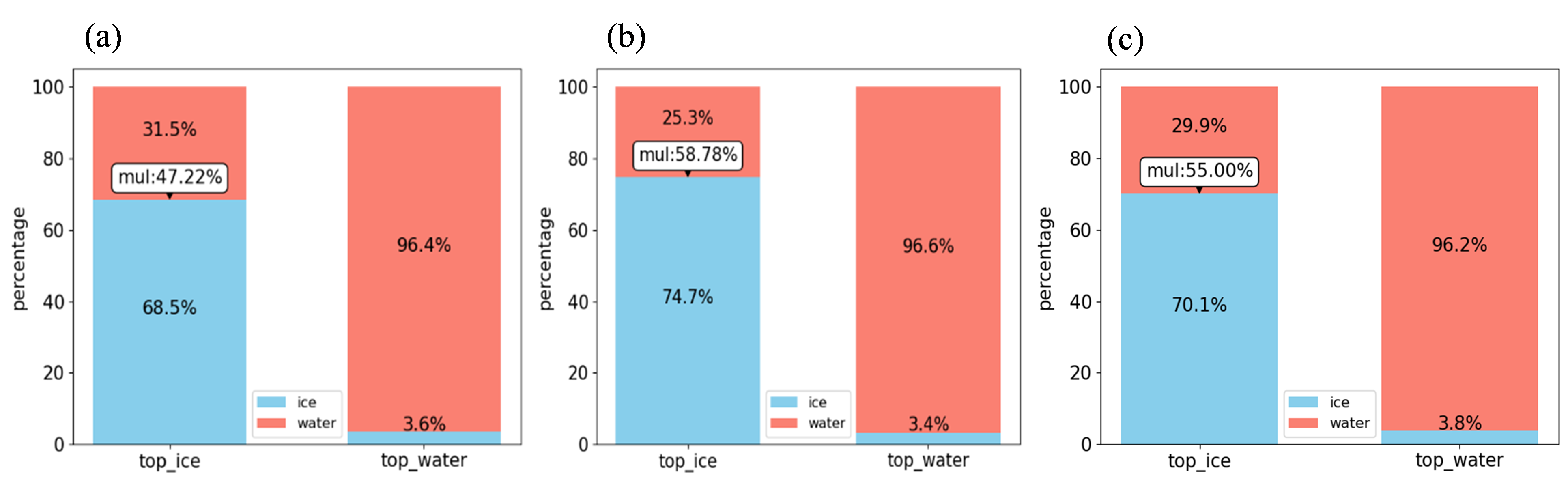}
    \caption{Comparison of the threshold-based algorithms' top-layer cloud phase identification performance against the CPR-CALIOP merged cloud product. (a) Baseline algorithm. (b) SubC14 threshold algorithm. (c) SubC13 threshold algorithm. Mul denotes the proportion of multilayer clouds with a correctly identified top phase.}
    \label{Fig 10}
\end{figure*}

\begin{figure*}[!t]
  \centering
  \includegraphics[width=\textwidth]{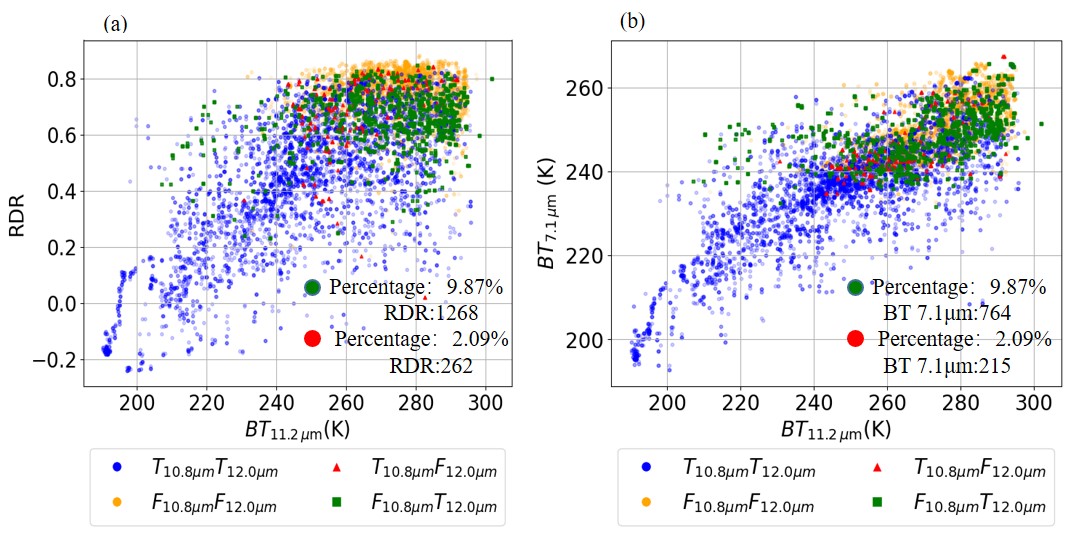}
  \caption{Threshold-based detection of top-layer cloud phase in multilayer clouds using the $10.8\,\mu$m baseline and $12.0\,\mu$m SubC13 threshold algorithms (T/F: correct/incorrect). (a) Detection results as a function of cloud optical depth and relative difference ratio (RDR). (b) Detection rates and proportions of samples within threshold bounds. (c) Detection results as a function of cloud optical depth and C10 brightness temperature.}
  \label{Fig 11}
\end{figure*}

\begin{figure*}[!t]
  \centering
  \includegraphics[width=\textwidth]{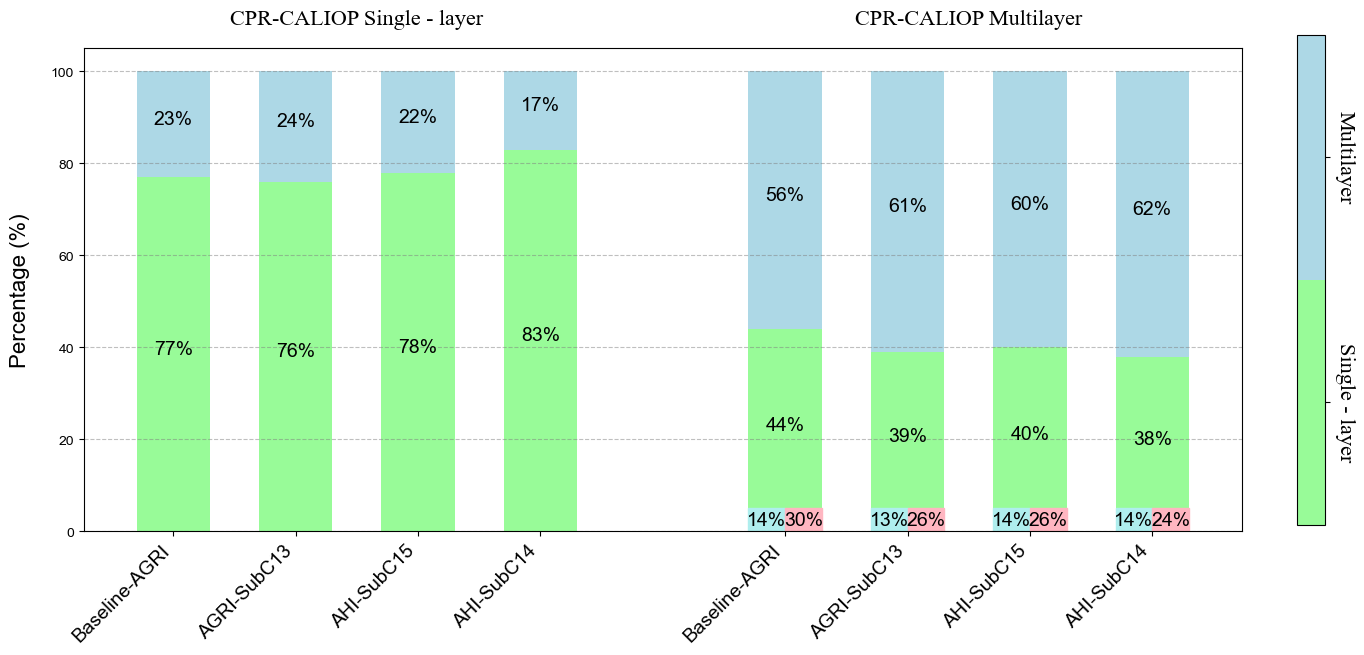}
  \caption{Percentages of single-layer and multilayer clouds detected by threshold-based algorithms with varied parameter configurations against CPR–CALIOP reference labels. Percentages of multilayer clouds misclassified as single-layer ice (light cyan) and water clouds (pale pink) are also shown.}
  \label{Fig 12}
\end{figure*}

Fig.~\ref{Fig 8} presents POD$_{\mathrm{mul}}$ and FAR$_{\mathrm{mul}}$ as functions of cloud optical thickness (COT) for the threshold-based experiment. Both SubC13 and the Baseline algorithm maintain FAR$_{\mathrm{mul}}\approx0.35$ across COT intervals, while SubC13 yields a consistently higher POD$_{\mathrm{mul}}$ (mean \(\approx0.6\)). The AHI experiments (AHI-SubC14 and AHI-SubC15) exhibit similar POD$_{\mathrm{mul}}$ \(\approx0.6\), but AHI-SubC15 shows a systematically larger FAR$_{\mathrm{mul}}$ than AHI-SubC14 at all COT levels. These patterns indicate that, despite apparent channel equivalence, instrument-specific channel configurations and spectral-response differences materially affect multilayer detection performance and therefore motivate adaptive, sensor-aware tuning of threshold parameters.

Fig.~\ref{Fig 9} shows that replacing C12 with C13 shifts the BT–BTD decision boundary, moving a substantial subset of multilayer samples into regions where $\mathrm{BT}_{12.0}$ and $\mathrm{BTD}_{8.5-12.0}$ provide a clearer separation of ice-top scenes than $\mathrm{BT}_{10.8}$ and $\mathrm{BTD}_{8.5-10.8}$. This boundary shift explains the improved top-layer ice identification and higher multilayer capture rate for the AGRI-SubC13 experiment, a conclusion that is quantitatively corroborated by the top-layer phase identification rates reported in Fig.~\ref{Fig 10}. As detailed in Fig.~\ref{Fig 9}, 47.22\% of top-layer ice in multilayer condition was correctly identified Fig.~\ref{Fig 9}(a)), but the correct rate reaches 55\% in the AGRI-SubC13 experiment (Fig.~\ref{Fig 9}(c)), second to AHI-SubC14 with the correct rate of 58.78\% (Fig.~\ref{Fig 9}(b)). Overall, substituting AGRI C13 for C12 increases correct top-layer ice identification by approximately 8 percentage points relative to the baseline algorithm, which helps explain the improved multilayer detection exhibited by the AGRI-SubC13 experiment.

Differences in branch-wise distributions can therefore reveal the relative capacity of AGRI C13 versus C12 to support multilayer detection. Fig.~\ref{Fig 11}(a) shows the two-dimensional joint distribution of top-layer phase identification outcomes for C12$_\mathrm{AGRI}$ and C13$_\mathrm{AGRI}$ in the BT$_{11.2\,\mu\mathrm{m}}$–RDR space: samples correctly classified by both channels are blue, those misclassified by both are orange, samples misclassified by C12 but correctly identified by C13 are green, and the converse case is red. Within the RDR (ice-top) threshold, 1{,}268 green multilayer samples were correctly recovered versus 262 red samples. Fig.~\ref{Fig 11}(b) shows the analogous distribution in the BT$_{11.2\,\mu\mathrm{m}}$–BT$_{7.1\,\mu\mathrm{m}}$ space: inside the BT$_{7.1\,\mu\mathrm{m}}$ threshold, 764 green multilayer samples and 215 red samples were correctly recovered. Hence, SubC13 identifies 1{,}268 (ice-top) + 215 (water-top) = 1{,}483 multilayer samples in these branches, while the Baseline identifies 262 + 764 = 1{,}026 samples under the same conditions.

Fig.~\ref{Fig 12} compares threshold-based multilayer cloud detection with the active reference (CPR–CALIPSO), focusing on single-layer versus multilayer classification after excluding pixels labeled as transparent. The evaluation only includes pixels with total cloud optical thickness (COT) $>5$. In the ``CPR–CALIPSO single-layer'' scenario, the SubC13, SubC15, and baseline threshold algorithms achieve comparable single-layer detection accuracies (~77\%), while the SubC14 threshold algorithm yields the best performance (~83\%). In the ``CPR–CALIPSO multilayer'' scenario, misclassifications into single-layer ice clouds and single-layer water clouds are reported. Compared with the baseline algorithm, the SubC13 threshold algorithm reduces the misclassification rates of single-layer ice and single-layer water clouds by 1\% and 4\%, respectively. These comparisons suggest that AGRI C13 offers a slight advantage over C12 in this classification task, highlighting the benefit of spectral-response optimization for threshold methods.

\begin{figure*}[htp]
    \centering
    \includegraphics[scale=1.2]{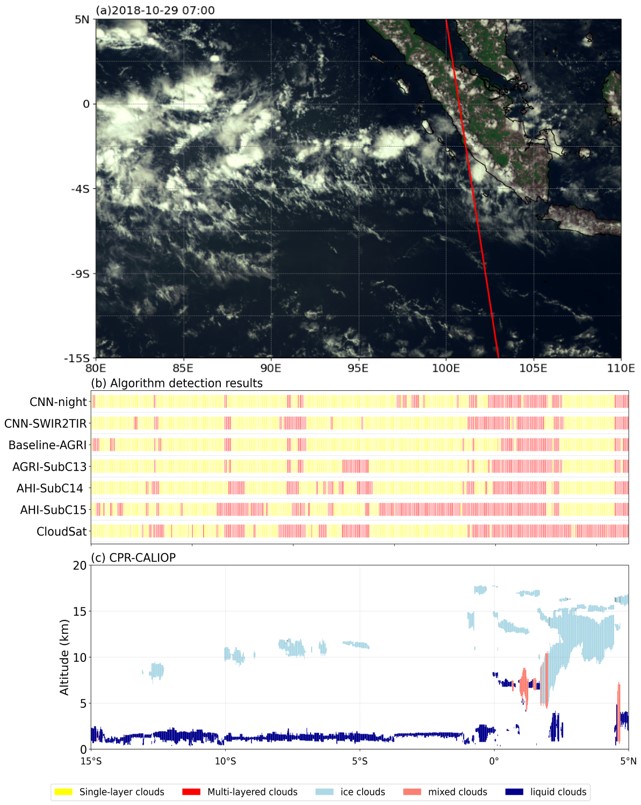}
    \caption{Case at 0700 UTC on 29 October 2018 over the East Asian continent. (a) True-color composite from Himawari-8 AHI channels (0.65\,$\mu$m, 0.86\,$\mu$m, and 0.46\,$\mu$m). (b) Cloud phase detection results using 1D-CNN models with varied parameter configurations and threshold-based algorithms. (c) Vertical cross-track profile of cloud thermodynamic phase from the CPR–CALIOP merged product along the orbit.}
    \label{Fig 13}
\end{figure*}

\subsection{Case Study over the Maritime Continent}

A representative case centered over the Maritime Continent (100°–105°E, 5°N–15°S) at 07:00 UTC on 29 October 2018 was selected to evaluate the cross-sensor consistency between algorithms, with CPR–CALIOP joint product serving as ground truth. Fig.~\ref{Fig 13}(a) shows an AHI natural-color composite (0.65, 0.86, 0.46\,$\mu$m) overlaid with the CPR–CALIOP ground track (red). Two distinct regimes are evident along the track: upper-level cirrus near the Japanese coast and low-level stratocumulus. Fig.~\ref{Fig 13}(b) compares detections from CNN models (notably CNN-SWIR2TIR and CNN-night) and threshold variants using the merged active-product labels; unlike the full-channel CNN-SWIR2TIR, CNN-night excludes near-infrared channels. The threshold comparison also assesses the effect of replacing {C12}$_\mathrm{AGRI}$ with {C13}$_\mathrm{AGRI}$ (SubC13), and the analogous substitution on AHI (SubC14). Fig.~\ref{Fig 13}(c) presents the vertical thermodynamic-phase profiles along the CPR–CALIOP track. For consistency, single-layer ice, water, and mixed-phase clouds are grouped as single-layer, and clear-sky pixels are excluded.

Overall, CNN-SWIR2TIR exhibits stronger agreement with the active-sensor reference than the threshold methods. In contrast, CNN-night shows pronounced false detections near $-15^\circ$S and $-2^\circ$, attributable to the absence of near-infrared channels, which leads to confusion of single-layer water clouds with multilayer clouds. For the threshold methods, SubC13 correctly identifies multilayered clouds near $-4^\circ$S that the baseline algorithm classifies as single-layer, consistent with the higher sensitivity of {C13}$_\mathrm{AGRI}$ to thin upper ice in multilayer systems. Similar behavior is observed for the AHI-based SubC14 variant.

\section*{Conclusion}
This study developed a framework that employs prior physical knowledge to guide feature engineering, combined with a 1D‑CNN for feature extraction, for multilayer cloud detection using geostationary imagers such as FY‑4A/AGRI and Himawari‑8/AHI. By embedding threshold-derived channel selections as physical priors, the network effectively bridges traditional algorithms and data-driven learning, achieving a POD$_\mathrm{mul}$ of~0.620 and a FAR$_\mathrm{mul}$ of~0.240, significantly surpassing the baseline threshold method. Gradient-based analysis further identifies the 12.0\,$\mu$m channel as the most informative predictor, whose inclusion consistently enhances both CNN- and threshold-based retrieval performance. Cross-sensor evaluations highlight that radiometric stability and spectral-response differences remain key factors affecting model generalization and operational transfer.

In future work, this physics–machine learning synergy will be extended to longer time periods and global-scale studies to further validate the proposed paradigm across diverse seasons, surface types, and viewing geometries.

\bibliographystyle{IEEEtran} 
\bibliography{bare_jrnl_new_sample4-bibliography}
\end{document}